\documentclass[a4paper]{spie}  
\usepackage[]{graphicx}

\title{TES imaging array technology for \CLOVER} 

\newcommand{\Clover}{{
  C\kern-.0667em\lower.6ex\hbox{$\ell$}\kern-.025emOVER}}

\newcommand{\CLOVER}{{
  C\kern-.0667em\lower.6ex\hbox{\Large$\ell$\normalsize}\kern-.025emOVER}}

\def\plotfiddle#1#2#3#4#5#6#7{\centering \leavevmode
\vbox to#2{\rule{0pt}{#2}}
    \includegraphics{#1}}

\author{Michael D. Audley, Robert W. Barker, Michael Crane, Roger Dace, Dorota Glowacka, David J. Goldie, Anthony N. Lasenby, Howard M. Stevenson, Vassilka Tsaneva, Stafford Withington\supit{a}, Paul Grimes, Bradley Johnson, Ghassan Yassin\supit{b}, Lucio Piccirillo, Giampaolo Pisano\supit{c}, William D. Duncan, Gene C. Hilton, Kent D. Irwin, Carl D. Reintsema\supit{d}, Mark Halpern\supit{e}
\skiplinehalf
\supit{a}Cavendish Laboratory, University of Cambridge, 
JJ Thomson Avenue, 
Cambridge, CB3 0HE, UK; \\
\supit{b}Department of Physics, University of Oxford, Denys Wilkinson Building, Keble Road, Oxford, OX1 3RH, UK;\\
\supit{c}School of Physics and Astronomy,
	Cardiff University,
	5, The Parade,
	Cardiff,
	CF24 3YB,
	Wales, UK.; \\
\supit{d}National Institute of Standards and Technology, 325 Broadway, Boulder, CO 80305, USA; \\
\supit{e} University of  British  Columbia, Department of Physics and Astronomy, 6224 Agricultural Rd., Vancouver, B.C., V6T 1Z1, Canada
}

\authorinfo{Further author information: (Send correspondence to M.D.A.)\\M.D.A.: E-mail: audley@mrao.cam.ac.uk, Telephone: +44(0)1223 337 309}

\begin{document} 
  \maketitle 

\begin{abstract}
\Clover\ is an experiment which aims to detect the signature of gravitational     
waves from inflation by measuring the B-mode polarization of the cosmic         
microwave background.  \Clover\ consists of three telescopes operating at 97, 150,
and 220 GHz.  The 97-GHz telescope has 160 horns in its focal plane while the   
150 and 220-GHz telescopes have 256 horns each.  The horns are arranged in a    
hexagonal array and feed a polarimeter which uses finline-coupled TES bolometers
as detectors.  To detect the two polarizations the 97-GHz telescope has 320     
detectors while the 150 and 220-GHz telescopes have 512 detectors each. To      
achieve the required NEPs the detectors are cooled to 100 mK for the 97 and     
150-GHz polarimeters and 230 mK for the 220-GHz polarimeter.  Each detector is  
fabricated as a single chip to guarantee fully functioning focal planes.  The detectors are contained  
in linear modules made of copper which form split-block waveguides.  The        
detector modules contain 16 or 20 detectors each for compatibility with the     
hexagonal arrays of horns in the telescopes' focal planes.  Each detector module
contains a time-division SQUID multiplexer to read out the detectors.  Further  
amplification of the multiplexed signals is provided by SQUID series arrays.    
The first prototype detectors for \Clover\ operate with a bath temperature of 230 
mK and are used to validate the detector design as well as the polarimeter technology.  We describe the design of the \Clover\ detectors, detector blocks,   
and readout, and give an update on the detector development.\footnote{Copyright 2006 Society of Photo-Optical Instrumentation Engineers.
This paper will be published in Proceedings of SPIE vol.~6275 and is made available
as an electronic preprint with permission of SPIE. One print or electronic copy may be made for personal
use only. Systematic or multiple reproduction, distribution to multiple locations via electronic or other means, duplication
of any material in this paper for a fee or for commercial purposes, or modification of the content of the paper
are prohibited.}

\end{abstract}



\section{INTRODUCTION}
\label{sect:intro}  

\subsection{Scientific motivation}
\label{sect:motivation}
Thomson scattering of primaeval radiation in the early Universe can lead to linear polarization\cite{Rees68} in the cosmic microwave background (CMB).  The polarization depends on density fluctuations, and thus carries cosmological information which is complimentary to the well-studied temperature anisotropies of the CMB.  The linear polarization may be decomposed into a curl-free part and a divergence-free part, denoted E- and B-mode respectively, by analogy with the electric field strength $E$ and magnetic induction $B$.  Linear density perturbations do not produce B-mode polarization, while tensor perturbations, as might be produced by gravitational waves, produce E- and B-mode polarization with similar amplitude\cite{Selj97,Kami97}.  Thus, by measuring the B-mode polarization of the CMB with \Clover\ we hope to make an indirect detection of a background of primordial gravitational waves.

\subsection{Overview of \Clover}
\label{sect:overview}
The \Clover\ experiment is described in detail elsewhere in these proceedings\cite{Call06}.
\Clover\ consists of three telescopes observing at frequencies of 97, 150, and 220~GHz.  The focal plane of each telescope will be populated by feed horns, each connected to a polarimeter.  The polarimeter technology is yet to be determined.  As part of the technology development program for \Clover\ we are constructing a Single Pixel Demonstrator.  This instrument will contain six TES detectors, allowing us to evaluate different polarization technologies, as well as validating the detector design.  

\section{\Clover\ DETECTORS} 
\label{sect:detectors}
\subsection{Detector Requirements}
\label{sect:requirements}
For maximum sensitivity, we require that the detectors be background-limited, i.e. the contributions to the noise equivalent power (NEP) from the detectors and readout must be less than half of the photon noise from the sky:
\begin{equation}
NEP_{det}^2+NEP_{ro}^2\le{1\over4}NEP_{photon}^2
\end{equation}
Once the detectors are background-limited the only way to improve the sensitivity is to increase the number of detectors.  \Clover's sensitivity requirements mean that we need 160 horns at 97~GHz and 256 each at 150 and 220~GHz.  Because the polarimeter splits the power from each horn into two modes which must be measured independently, the number of detectors is twice this.  We require a detector time constant of less than 1~ms.
Also, the detectors must be able to absorb the power incident from the sky without saturation.  This power is variable and depends on the weather.  The power-handling requirement is for the detectors to be able to operate for 75\%\ of the time at the site.  The detector requirements are summarized in Table~\ref{tab:requirements}.

\begin{table}[h]
\caption{\Clover\ detector requirements at the three operating frequencies.} 
\label{tab:requirements}
\begin{center}       
\begin{tabular}{|c|c|c|c|c|} 
\hline
\rule[-1ex]{0pt}{3.5ex}  Centre Frequency &Band&Number of detectors&NEP Requirement&Power Handling\\
\rule[-1ex]{0pt}{3.5ex}  (GHz)&(GHz)&&($10^{-17}\rm\ W/\sqrt Hz$)&(pW)  \\
\hline
\rule[-1ex]{0pt}{3.5ex}  97& 82--112&320&1.5& 6.7\\
\hline
\rule[-1ex]{0pt}{3.5ex}  150& 127.5--172.5&512&2.5& 11.5\\
\hline
\rule[-1ex]{0pt}{3.5ex}  220 & 195--255&512&4.5&  18.8\\
\hline 
\end{tabular}
\end{center}
\end{table} 

\subsection{Detector Architecture} 
The detector architecture has changed since the original concept.  Originally, \Clover\ had four identical telescopes at each frequency.  The detector chips would contain eight TESs, each of which would be fed by four finlines.  Each of these four finlines would receive the same mode from one of the four telescopes, improving the sensitivity of each detector.  This meant that the number of detectors would be a quarter of the number of modes from the polarimeters.  This scheme was dropped in favour of having a single detector per mode, not least because of the difficulty of plumbing such a complicated waveguide configuration.  The current configuration has a single TES on each chip, fed by a single finline.  There are two main reasons for going against the current trend towards large monolithic arrays.  First, because the focal plane is populated by feedhorns, the waveguides coming from the polarimeter are on a pitch which is much larger than the size of the detectors.  The horn diameters are 20, 8.4, and 12.77~mm at 97, 150, and 220~GHz, respectively.  A monolithic array would have large inactive areas between active elements, increasing the number of wafers that would have to be processed, and hence the cost and manufacturing time.  Second, because high sensitivity is essential for achieving \Clover's science goals, we decided to fabricate each detector on a single chip so that good devices could be selected to guarantee that all the detectors in each focal plane are working.  The chips are micromachined and diced by deep reactive ion etching (DRIE) at the Scottish Microelectronics Centre.  

Another change to the instrument is that \Clover\ will now be sited in the Atacama Desert in Chile, instead of Dome~C in Antarctica, as was originally planned.  This affects the detector power-handling and sensitivity requirements (see Sect.~\ref{sect:requirements}).  The main features of the \Clover\ detectors and how they have changed are summarized in Table~\ref{tab:changes}.

\begin{table}[h]
\caption{Current \Clover\ detector architecture compared with original concept.} 
\label{tab:changes}
\begin{center}       
\begin{tabular}{|c|c|c|} 
\hline
\rule[-1ex]{0pt}{3.5ex}Feature&Original&Current\\
\hline
\rule[-1ex]{0pt}{3.5ex}  detectors per chip& 8& 1\\
\hline
\rule[-1ex]{0pt}{3.5ex}  microstrips per detector& 4& 1\\
\hline
\rule[-1ex]{0pt}{3.5ex}  Multiplexing& Frequency-domain & Time-domain \\
\hline
\rule[-1ex]{0pt}{3.5ex}  Multiplexing level & 8& 16 (97 GHz)\\
\rule[-1ex]{0pt}{3.5ex}  & &20 (150, 220 GHz)\\
\hline 
\end{tabular}
\end{center}
\end{table}

\subsection{R.F. Design} 
\label{sect:rfdesign}
\begin{figure}
   \begin{center}
   \begin{tabular}{c}
   \includegraphics[height=7cm]{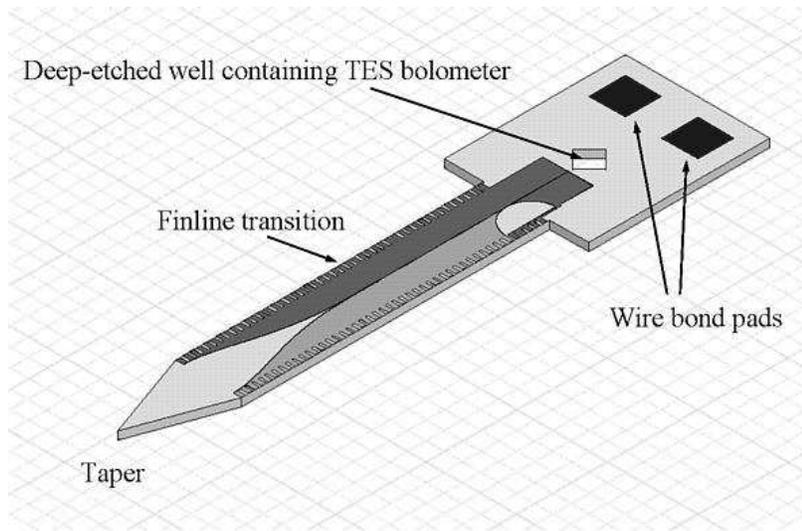}
   \end{tabular}
   \end{center}
   \caption[chip] 
   { \label{fig:chip} 
Layout of prototype \Clover\ detector chip.}
   \end{figure}

To reach background-limited sensitivity \Clover's bolometers must have a high absorption efficiency. Power is coupled from the waveguide to the TES planar circuit using an antipodal finline taper consisting of two superconducting fins of Nb separated by 400~nm of SiO$_2$\cite{Yass95,Yass2000} (see Fig.~\ref{fig:chip}). The lower Nb layer is 250~nm thick.  The upper layer is 500~nm thick to ensure reliable lift-off patterning with the step over the oxide layer.  The whole structure is deposited on one side of a 225-$\mu\rm m$ silicon substrate. Before the fins overlap, the thickness of the SiO is much less than that of the silicon and the structure behaves as a unilateral finline. As the fins overlap, the structure starts to behave like a parallel-plate waveguide with an effective width equal to the overlap region. When the width of the overlap region becomes large enough for fringing effects to be negligible, a transition to a microstrip mode has been performed. The microstrip is then tapered to the required width.  \Clover\ uses a 3-$\rm\mu m$ Nb microstrip with a characteristic impedance of $20\rm\ \Omega$ to deliver power to the TES.

The detector chip's 225-$\rm\mu m$ silicon substrate loads the waveguide in which it sits, changing the waveguide impedance.  To prevent reflections the chip has a tapered end which provides a gradual impedance transition.  The prototype detectors for the Single Pixel Demonstrator sit in a WR-10 waveguide and have a taper angle of $40^\circ$.  This angle was chosen based on finite-element electromagnetic modelling.

\subsection{Bolometer Design}
\begin{figure}
   \begin{center}
   \begin{tabular}{c}
   \includegraphics[height=7cm]{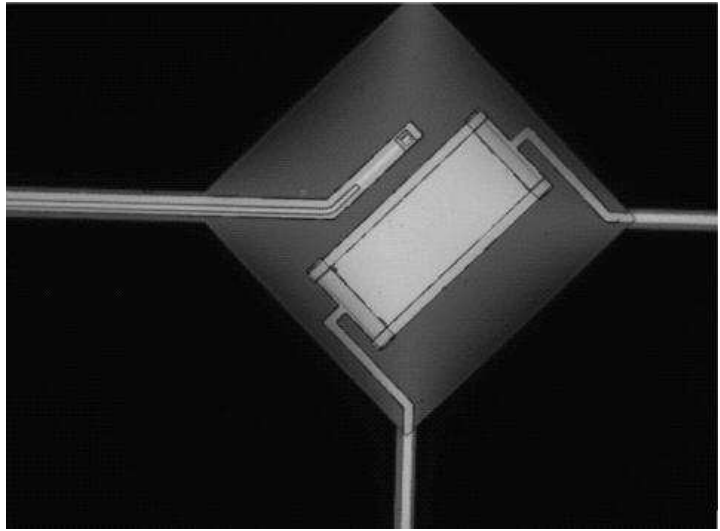}
   \end{tabular}
   \end{center}
   \caption[island] 
   { \label{fig:island} 
\Clover\ prototype bolometer silicon nitride island showing TES and microstrip leading to termination resistor.}
   \end{figure}

\Clover's bolometers are low-stress silicon nitride islands suspended on four legs (see Fig.~\ref{fig:island}).  The nitride is 0.5~$\rm\mu m$ thick.  The thermal conductance to the thermal bath is controlled by the four nitride legs.  The microstrip carrying power from the finline to the bolometer is terminated by a 20-$\Omega$ Au/Cu resistor which dissipates the incoming power as heat that the superconducting transition edge sensor (TES) can detect.  A shunt resistor in parallel with the TES ensures that it is voltage biased so that it operates in the regime of strong negative electrothermal feedback\cite{KentsThesis}. For example, if the temperature drops, so does the resistance of the TES. Since it is biased at constant voltage, this means that the current, and hence the Joule power, will increase, heating up the TES. Conversely, if the temperature increases the resistance will increase, reducing the current, and thus the Joule heating. This means that the TES operates at a bias point that is in a stable equilibrium. Thus, the TES is self-biasing. There is no need for a temperature controller to ensure that it remains at the correct bias point. Also, the electrothermal feedback cancels out temperature fluctuations which has the effect of suppressing the Johnson noise. 

\begin{figure}
\plotfiddle{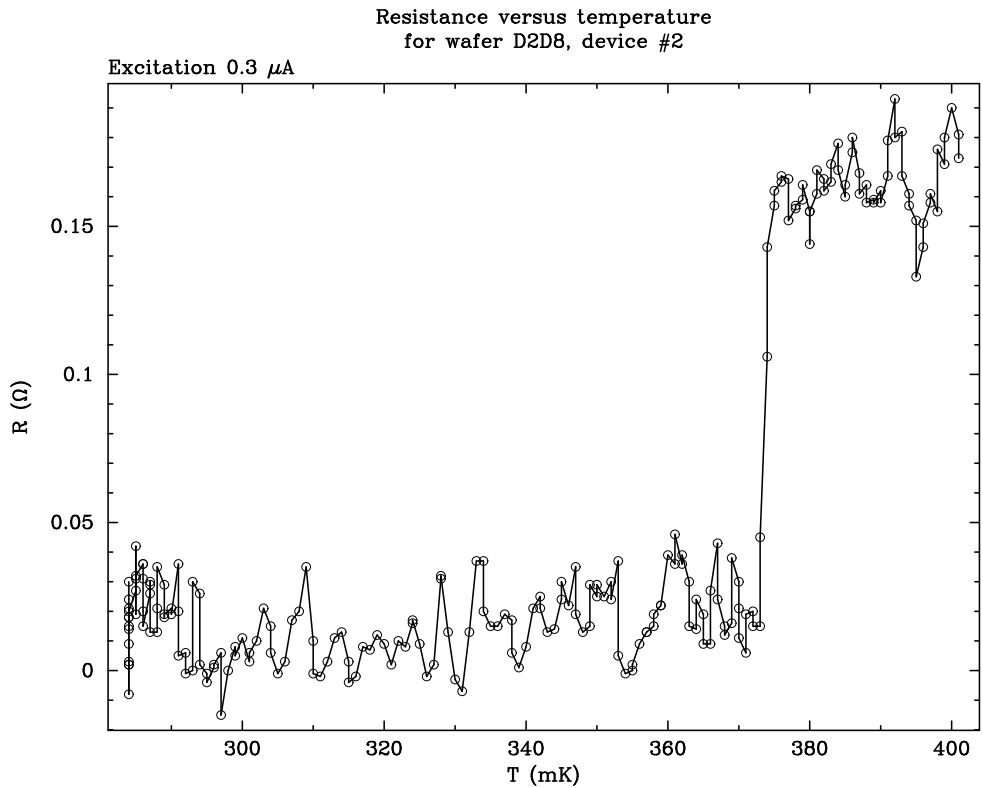}{293.874pt}{270}{54.3}{54.3}{-196.640pt}{309.627pt}
   \caption[RvsT] 
   { \label{fig:RvsT} 
Resistance versus temperature plot for one of the prototype \Clover\ detectors.  The transition is 1--2~mK wide and $T_c=373.5\rm\ mK$.}
   \end{figure}

The TES films in \Clover\ are Mo/Cu proximity-effect bilayers. The transitions of the bilayers can be made as sharp as 1--2 mK for high sensitivity.  The sensitivity of the TES shown in Fig.~\ref{fig:RvsT} is $\alpha={T\over R}{dR\over dT} = \frac{d\log R}{d\log T}\approx300$.  We can also tune the transition temperature ($T_c$) of the films to the desired value by choosing the film thicknesses.

The operating temperature of \Clover's detectors is chosen to meet the NEP requirements and is dominated by the phonon noise.
  Cooling is provided by a Cryomech PT-405 pulse-tube cooler, a high-capacity Simon Chase He-7 cooler, and a miniature dilution refrigerator\cite{Tele06}.  Because a TES is a low-impedance device it is not very susceptible to microphonics, making it feasible to use a pulse-tube cooler.  The 97 and 150-GHz detectors will operate with a base temperature of 100~mK and $T_c=190\rm\ mK$, while the 220-GHz detectors require a base temperature of 230~mK and $T_c=430\rm\ mK$.

\begin{figure}
\plotfiddle{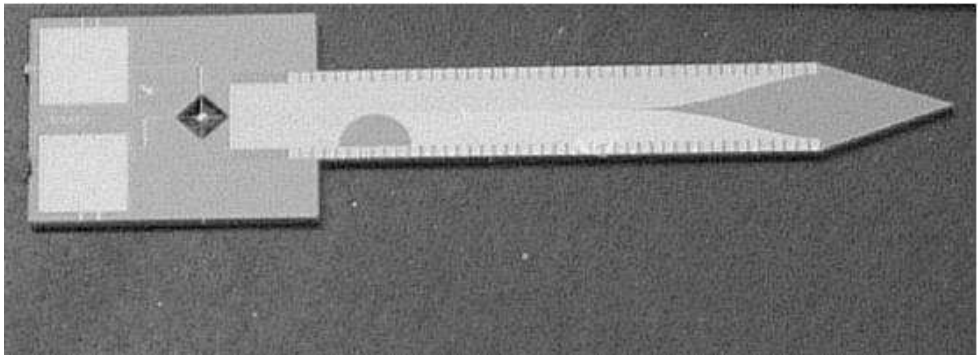}{132.173pt}{0}{66.1}{66.1}{-183.060pt}{0.000pt}
   \caption[tweezers] 
   { \label{fig:tweezers} 
Prototype \Clover\ detector chip.  The chip is about 16~mm long.}
   \end{figure}

All of the detectors for the Single Pixel Demonstrator have been fabricated and are undergoing testing.  Figure~\ref{fig:tweezers} shows a prototype \Clover\ detector chip.  We plan to carry out RF measurements of the \Clover\ detectors in the near future using a cryogenic black body source.

\section{READOUT} 
\begin{figure}
   \vspace{9pt}
   \begin{center}
   \begin{tabular}{c}
   \includegraphics[height=7cm]{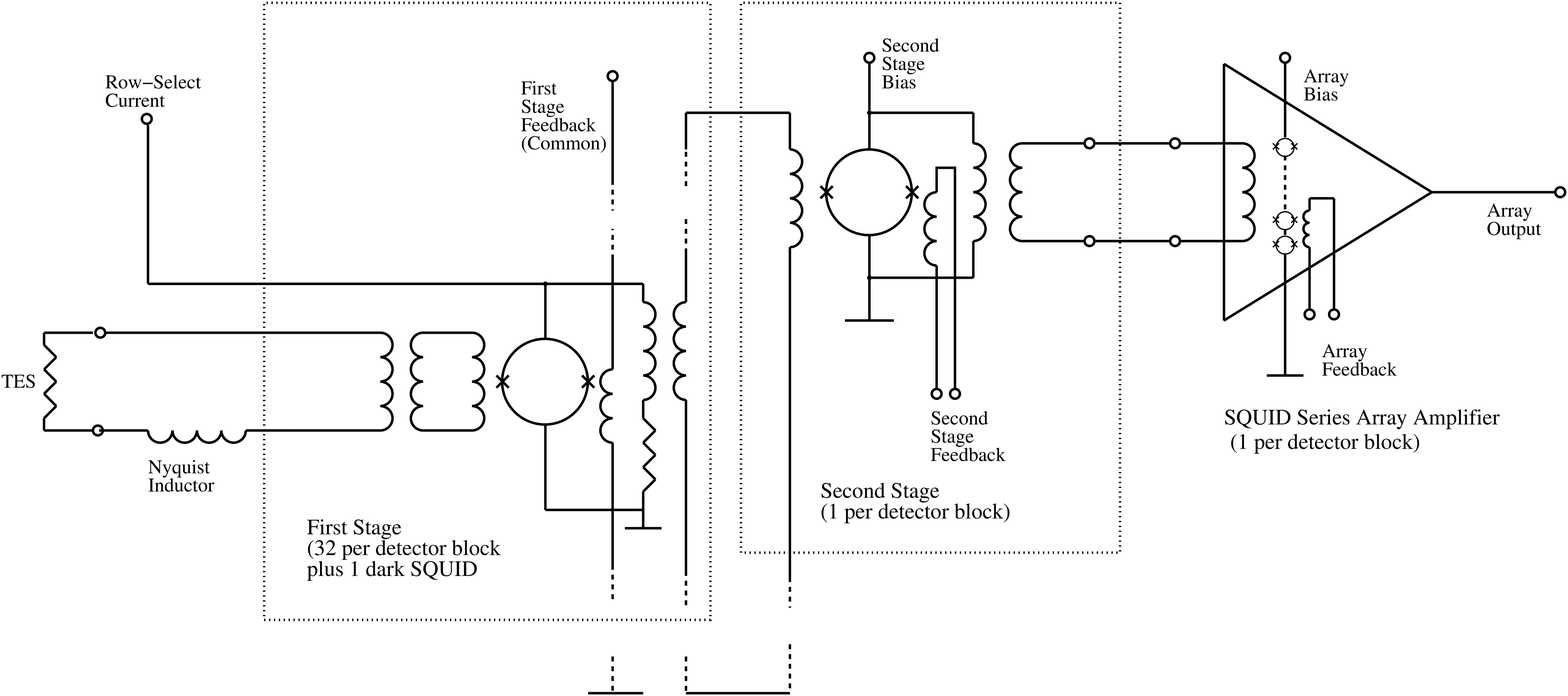}
   \end{tabular}
   \end{center}
   \caption[circuit] 
   { \label{fig:circuit} 
Simplified diagram of the \Clover\ readout.  The dummy SQUID that cancels out crosstalk is not shown.  Note that all grounding is done outside the cryostat.  For every line that enters the instrument there is a return line.}
   \end{figure}

Given the large number of detectors in this instrument (320 at 97~GHz and 512 each at 150 and 220~GHz) some form of multiplexing is needed to have a manageable number of wires from room temperature.  We use $1\times32$ SQUID multiplexers\cite{Cher99,deKorte03} fabricated by the National Institute of Standards and Technology.  A simplified view of the \Clover\ readout is shown in Fig.~\ref{fig:circuit}.  The first and second stage SQUID amplification stages are contained in a $1\times32$ multiplexer chip.  The first stage is actually a balanced pair of SQUIDs. One of the SQUIDs in a pair is unbiased and cancels out crosstalk in unaddressed pixels as the SQUID
bias and feedback are switched from row to row. There is also a 33$^{\mbox{rd}}$ dark SQUID on each chip which is
used to cancel out drift ($1/f$ noise) in the downstream electronics.  All the multiplexer chips in each of \Clover's three telescopes share address lines, significantly reducing the number of wires needed to room temperature.  The Nyquist inductors, which slow down the circuit to make it stable against oscillations, are contained in a seperate chip.  Both of these chips and the shunt resistors are mounted on a PCB at 100~mK and connections are made to the detectors by aluminium wire bonds.  The SQUID series arrays\cite{Welt93} that provide the third stage of amplification are mounted in eight-chip modules which provide the necessary magnetic shielding.  These modules are heat-sunk to the 4-K stage of the cryostat and they are connected to the multiplexer PCB with superconducting NbTi twisted pairs.  Room-temperature multi-channel electronics (MCE), developed by the University of British Columbia, provide SQUID control and readout as well as TES bias.  \Clover's MCE is 
similar to that used by SCUBA-2\cite{SPIE04}. 

\section{DETECTOR Packaging} \label{sect:sections}

\subsection{Populating the Focal Plane} 
\begin{figure}
   \begin{center}
   \begin{tabular}{c}
   \includegraphics[height=7cm]{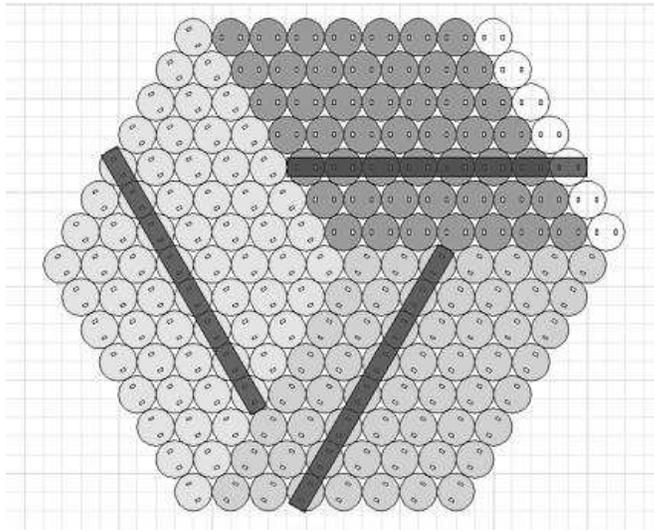}
   \end{tabular}
   \end{center}
   \caption[hexeight] 
   { \label{fig:hexeight} 
Layout of 90-GHz focal plane.  The dark rectangles show the orientation of linear detector blocks.}
   \end{figure}

The feedhorns are arranged in a hexagonal array.  However, the $1\times32$ multiplexer chips we are using lend themselves more naturally to a planar configuration where we have up to 16 horns in a row.  We therefore had to come up with a compromise arrangement which would allow us to tile the focal plane efficiently.  As shown in Fig.~\ref{fig:hexeight} we split the 97-GHz focal plane into three regions.  Clockwise from upper right these are a $7\times7$-horn parallelogram, an $8\times7$-horn parallelogram, and an $8\times8$-horn parallelogram.  The two waveguides corresponding to each horn are arranged so that they are all parallel within one of these regions.  This allows us to cover each region with linear detector blocks stacked on top of each other with an offset to match the hexagonal horn pitch.  The orientation of one of these detector blocks is shown by a dark rectangle in each of the three regions.  The main advantage of this arrangement is that it allows us to use identical detector blocks to cover a focal plane, reducing cost and complexity.  The 97-GHz focal plane needs 22 detector blocks to cover it.  The scheme for covering the 150- and 220-GHz focal planes is similar, except that the horns are arranged in a hexagon with a side of ten horns.  This means that there are 28 detector blocks, each containing 20 detectors and one $1\times32$ multiplexer chip.  

This scheme has the apparent disadvantage that we are under-using the $1\times32$ multiplexer chips by a factor of up to two, increasing the number needed.  However, because we are not using all of the first-stage SQUIDs on a multiplexer chip, we can connect the detectors to those SQUIDs that have the most similar critical currents.  This optimizes the first-stage SQUID biasing, reducing the noise contribution from this stage of the readout.  Reducing the number of detectors multiplexed by each multiplexer chip also reduces the aliased readout noise, improving the NEP.  Another advantage of under-using the multiplexer chips is that we can use chips where not all the first-stage SQUIDs are functioning, reducing the cost per chip.

\subsection{Detector Block}
\begin{figure}
   \begin{center}
   \begin{tabular}{c}
   \includegraphics[height=7cm]{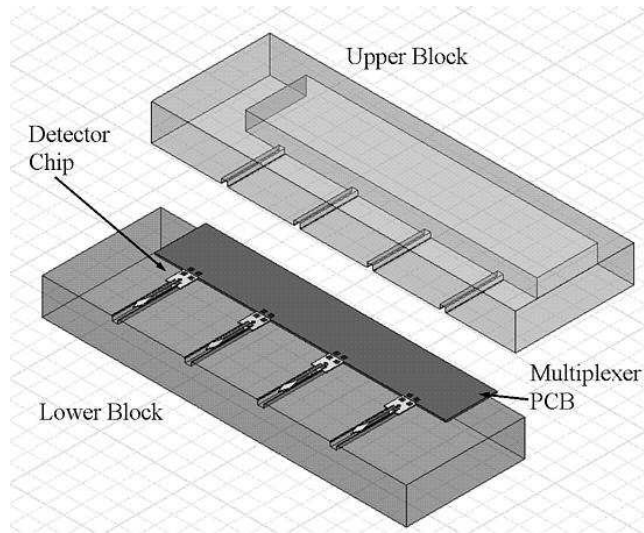}
   \end{tabular}
   \end{center}
   \caption[block] 
   { \label{fig:block} 
Detector block concept showing how four detectors would be mounted in a block.  The upper and lower blocks form waveguides in which the finlines sit.}
   \end{figure}

The detector block comes in two halves, upper and lower.  When these are put together they form split-block waveguides, into which the finlines protrude.  The edges of the finlines stick into shallow slots in the sides of the waveguides for grounding.  The serrations on the edges of the finlines (see Fig.~\ref{fig:chip}) are there to prevent unwanted modes from propagating.  A simplified view of a detector block holding four detectors is shown in Fig.~\ref{fig:block}.

Aluminium wire bonds provide electrical connections from the detector chip to a PCB carrying the multiplexer, inductors, and shunt resistors.  This PCB has gold-plated copper tracks and as much of the copper as possible is left on the board to help with heatsinking.  The gold is deposited by electroplating in order to avoid the use of a nickel undercoat.  The traces are tinned with solder to make them superconducting.  The PCB is enclosed in a copper can which is wrapped in niobium foil under which there is a layer of Metglas$^{\mbox{\textregistered}}$~2705M, a high-permeability amorphous metal foil (Hitachi Metals Inc.).  The Nb foil excludes magnetic fields while the Metglas diverts any trapped flux away from the SQUIDs.  Further magnetic shielding is provided by high-permeability shields built into the cryostat.

\subsection{Detector Mounting Scheme}
In the final instrument each detector block will carry either 16 or 20 detectors.  We would like to be able to remove and replace one of these detectors without disturbing the others.  Thus, we mount each detector chip on an individual copper chip holder, which is then mointed in the detector block.
 
We must make good thermal contact to the back of each detector chip, while at the same time relieving stresses caused by differential contraction that could demount or break the chip.  We fix the chip to a chip holder using epoxy.  We carried out thermal cycling tests with various different epoxies and the best results were obtained with Stycast 1266.  The chip holder has a well in the centre to divert excess epoxy away from the suspended nitride island.

The chip holder is secured to the detector block by two brass screws.
A piece of Nb foil between the chip holder and the detector block provides magnetic shielding for the detector chip.  


\acknowledgments     
 
\Clover\ is funded by the Particle Physics and Astronomy Research Council.
We are grateful to Andrew Bunting at the Scottish Microelectronics Centre of the University of Edinburgh for carrying out the DRIE and for his helpful advice.

\bibliography{../report_corrected}   

\begin{thebibliography}{10}

\bibitem{Rees68}
M.~J. {Rees}, ``{Polarization and Spectrum of the Primeval Radiation in an
  Anisotropic Universe},'' {\em \apjl} {\bf 153}, pp.~L1+, July 1968.

\bibitem{Selj97}
U.~{Seljak} and M.~{Zaldarriaga}, ``{Signature of Gravity Waves in the
  Polarization of the Microwave Background},'' {\em Physical Review Letters}
  {\bf 78}, pp.~2054--2057, Mar. 1997.

\bibitem{Kami97}
M.~{Kamionkowski}, A.~{Kosowsky}, and A.~{Stebbins}, ``{A Probe of Primordial
  Gravity Waves and Vorticity},'' {\em Physical Review Letters} {\bf 78},
  pp.~2058--2061, Mar. 1997.

\bibitem{Call06}
P.~G. Calisse, P.~A.~R. Ade, M.~D. Audley, P.~Cabella, A.~D. Challinor,
  P.~Ferreira, W.~K. Gear, D.~J. Goldie, P.~K. Grimes, K.~Isaac, M.~E. Jones,
  B.~Kiernan, M.~Crane, R.~W. Barker, K.~Grainge, P.~F. Horner, B.~J. Johnson,
  A.~N. Lasenby, B.~Maffei, P.~D. Mauskopf, S.~J. Melhuish, C.~E. North,
  A.~Orlando, S.~M. Parsley, L.~Piccirillo, L.~Pietranera, O.~E. Mallie,
  G.~Pisano, G.~Savini, H.~M. Stevenson, A.~C. Taylor, G.~Teleberg, I.~Thomas,
  V.~N. Tsaneva, C.~E. Tucker, R.~Tucker, I.~Walker, S.~Wilcox, S.~Withington,
  G.~Yassin, L.~Dunlop, D.~Glowacka, D.~T. O'Dea, G.~Rocha, and A.~C. Readhead,
  ``Clover: a high-sensitivity polarization instrument for {CMB} {B}-mode
  observations.'' SPIE 6275, in press, 2006.

\bibitem{Yass95}
G.~{Yassin} and S.~{Withington}, ``{Electromagnetic models for superconducting
  millimetre-wave and sub-millimetre-wave microstrip transmission lines },''
  {\em Journal of Physics D Applied Physics} {\bf 28}, pp.~1983--1991, Sept.
  1995.

\bibitem{Yass2000}
G.~Yassin, S.~Withington, M.~Buffey, K.~Jacobs, and S.~Wulff, ``A 350-{G}hz
  {SIS} antipodal finline mixer,'' {\em IEEE Trans. on Microwave Theory and
  Techniques} {\bf 48}(4), pp.~662--669, 2000.

\bibitem{KentsThesis}
K.~D. Irwin, {\em Phonon-{M}ediated {P}article {D}etection {U}sing
  {S}uperconducting {T}ungsten {T}ransition-{E}dge {S}ensors}.
\newblock PhD thesis, Stanford University, Stanford, California, 1995.

\bibitem{Tele06}
G.~Teleberg and L.~Piccirillo, ``A miniature dilution refrigerator for
  sub-{K}elvin detector arrays.'' SPIE 6275, in press, 2006.

\bibitem{Cher99}
J.~A. {Chervenak}, K.~D. {Irwin}, E.~N. {Grossman}, J.~M. {Martinis}, C.~D.
  {Reintsema}, and M.~E. {Huber}, ``{Superconducting multiplexer for arrays of
  transition edge sensors},'' {\em Applied Physics Letters} {\bf 74},
  pp.~4043--4045, June 1999.

\bibitem{deKorte03}
P.~A.~J. {de Korte}, J.~{Beyer}, S.~{Deiker}, G.~C. {Hilton}, K.~D. {Irwin},
  M.~{Macintosh}, S.~W. {Nam}, C.~D. {Reintsema}, L.~R. {Vale}, and M.~E.
  {Huber}, ``{Time-division superconducting quantum interference device
  multiplexer for transition-edge sensors},'' {\em Review of Scientific
  Instruments} {\bf 74}, pp.~3807--3815, Aug. 2003.

\bibitem{Welt93}
R.~P. {Welty} and J.~M. {Martinis}, ``{Two-stage integrated SQUID amplifier
  with series array output},'' {\em IEEE Transactions on Applied
  Superconductivity} {\bf 3}, pp.~2605--2608, Mar. 1993.

\bibitem{SPIE04}
M.~D. {Audley}, W.~S. {Holland}, T.~{Hodson}, M.~{MacIntosh}, I.~{Robson},
  K.~D. {Irwin}, G.~{Hilton}, W.~D. {Duncan}, C.~{Reintsema}, A.~J. {Walton},
  W.~{Parkes}, P.~A.~R. {Ade}, I.~{Walker}, M.~{Fich}, J.~{Kycia},
  M.~{Halpern}, D.~A. {Naylor}, G.~{Mitchell}, and P.~{Bastien}, ``{An update
  on the SCUBA-2 project},'' in {\em Millimeter and Submillimeter Detectors for
  Astronomy II. Proceedings of the SPIE, Volume 5498},  J.~{Zmuidzinas}, W.~S.
  {Holland}, and S.~{Withington}, eds., pp.~63--77, Oct. 2004.

\end{thebibliography}
\bibliographystyle{spiebib}   

\end{document}